%Paper: cond-mat/9410102
%From: Andrea Crisanti <krisanti@dectar.roma1.infn.it>
%Date: Thu, 27 Oct 1994 10:42:19 +0100

%%%%%%%%%%%%%%%%%%%%%%%%%%%%%%%%%%%%%%%%%%%%%%%%%%%%%%%%%%%%%%%%%%%%
%%
%% TeX file
%% input.sty is appended at the end as uuencoded compressed 
%% figures are appended as uuencoded compressed tar 
%%
%% for problems crisanti@roma1.infn.it
%%
\input preprint.sty
%
% ------------------- local definitions
\catcode`\@=11       % WARNING: we use private TeX codes !!!!
\def\BBig#1{{\hbox{$\left#1 \vbox to 20.5pt{} \right. \n@space$}}}
\def\BBigg#1{{\hbox{$\left#1 \vbox to 23.5pt{} \right. \n@space$}}}
\catcode`\@=12

      % this is to have \bdi in $$ $$
\font\extsyten=lasy10
\textfont10=\extsyten
\mathchardef\Box="2A32
\def\\{\hfill\break}
 
\def\Lap#1{\widetilde{#1}}

\def\xi{x_i}

\def\gu{\alpha}
\def\gd{\beta}

\def\lc{l_{\rm c}}
\def\Lap#1{\widetilde{#1}}

% ------------------- end local definitions

\title{Sporadicity and synchronization in one-dimensional asymmetrically 
       coupled maps}
 
\author{F Cecconi, A Crisanti, M Falcioni\dag\ and A Vulpiani\dag}
 
\address{Dipartimento di Fisica, Universit\`a di Roma 
       ``La Sapienza'', P.le A. Moro 2, I--00185 Roma, Italy}
\address{\dag\ INFN, Sezione di Roma}
 
\shorttitle{chaos in extended systems}
 
 
\jnl{JPA}
 
\date
 
\beginabstract
A one-dimensional chain of sporadic maps with asymmetric nearest neighbour 
couplings is numerically studied. It is shown that in the region of strong 
asymmetry the system becomes spatially fully synchronized, even in the 
thermodinamic limit, while the Lyapunov exponent is zero. 
For weak asymmetry the synchronization is no more complete, and the 
Lyapunov exponent becomes positive. In addition one has a clear 
relation between temporal and spatial chaos, {\it i.e.}: a positive 
effective Lyapunov exponent corresponds to a lack of synchronization 
and {\it vice versa}
\endabstract

An interesting property of extended systems is that they can exhibit 
complex behaviour both in space and time, that is the 
chaotic evolution of a spatial pattern. This kind of phenomena, 
called ``spatio-temporal'' chaos, has been the focus of considerable 
interest in different fields, as chemical reaction-diffusion 
systems, B\'enard convection, turbulence, modelling of brain fuction, and 
so on [1-2]. Recent experiments have revealed the emergence of large-scale 
spatio-temporal patterns of activity in many brain areas, such as the 
olfactory system or the visual cortex [3]. 
Coherent spatial patterns play also an important role in the behaviour of 
turbulent fluids. Here 
numerical simulations and experiments show the emergence of strong
choerent vortex structures which are responsible for intermittency and 
possible anomalous dimensions in the scaling law [4]. Moreover choerent 
structures may play a non trivial role in the predictabilty problem based
on two-dimensional turbulence, an important subjec from both theoretical 
and metereological point of view [5]. 

A direct study of the spatio-temporal behaviour from the full equations, 
such as the Navier-Stokes equations, in general, is quite complicated, 
even numerically. To overcome this problem, and gain more insight in 
the basic mechanisms, much work has been 
devoted to the study of models, such as coupled lattice maps. 
These are a crude, but nontrivial, approximation of extended systems with 
discrete space and time, but continuous states [2]. The simplest form
is given by a 
set of $N$ continuous variables $x_i$ which evolve in (discrete) time as
   $$\eqalign{
              x_i(n+1)&= (1-\gu_i-\gd_i)\, f(x_i(n)) \cr
           &\phantom{=}
             + \gu_i f(x_{i-1}(n)) + \gd_i f(x_{i+1}(n)). \cr} 
    \eqno(1)$$
The index $i$ denotes the $i^{th}$ site on a one-dimensional lattice, and 
only nearest neighbour interactions are taken into account. 
The strenght of the couplings and their spatial homogeneity, as well as 
the type of boundary conditions used, may depend on the the type 
of physical problem. For example in the case of shear flow, boundary layers 
or convection, there is a privileged direction. This can be introduced in the 
model (1) by taking asymmetric couplings [6,7]. 

Moreover in such systems the most appropriate boundary conditions are
open boundary conditions. Thermal reservoir can be included by giving a 
special evolution law to the sites on the boundary. We do not consider 
this generalization here. 

In this letter we shall study a one-dimensional chain of 
coupled maps (1) with homogeneous asymmetric couplings and open boundary 
conditions. The equations of motion are given by (1), with couplings
   $$\eqalign{
              \gu_i=& \gamma_1,\quad  \gd_i=\gamma_2 \quad
                       \hbox{\rm for}\ i=2,\dots,N-1 \cr 
              \gu_1=& 0,\quad \gd_1= \gamma_2, \quad
              \gu_N=\gamma_1,\quad \gd_N=0.\cr}
    \eqno(2)$$
Without loss of generality we assume $\gamma_1 > \gamma_2$.
The system (1) with these couplings was studied in Refs. [7,8] for the 
case of chaotic single map $f(x)$. In these papers the interest was for the 
possible emergence of spatial coherent patterns due to the asymmetrical 
couplings. 

A simple inspection reveals that the system (1)-(2) posses the uniform solution
   $$x_i(n)=\Lap{x}(n), \qquad
     \Lap{x}(n+1)=f(\Lap{x}(n)).
    \eqno(3)$$ 
To study the stability of the uniform state one can linearize eq. (1) 
about the uniform solution (3) and study the spectrum of the 
fluctuations [7]. This consists of a uniform eigenmode with eigenvalue
$\rho_0$ and $N-1$ nonuniform eigenmodes with spectrum
   $$\rho(k) = \rho_0 \left[
                1 - \gamma_1 - \gamma_2 + 2\sqrt{\gamma_1\gamma_2}\,\cos(k)
                      \right]
    \eqno(4)$$
where $k = \pi m / N$ with $m = 1, 2,\ldots,N-1$, and 
$\rho_0= \exp(\lambda_0)$ where $\lambda_0$ is the largest Lyapunov exponent 
of the single map $f(x)$. 

For $\gamma_1\not=\gamma_2$ this spectrum posses a gap at $k=0$ since 
$\rho(k\to 0)$ is less that the $k=0$ eigenvalue $\rho_0$. 
Therefore if $\gamma_1$ and $\gamma_2$ are such that $\rho_0>1$ 
and $|\rho(k)|<1$ 
then all nonuniform fluctuations are stable. The only instability that is 
left is the instability to uniform fluctuations, which is inherent to the 
chaotic nature of the single map. 

>From this one concludes that the uniform state (3) is stable. This 
predictions for the case of a single chaotic map $f(x)$  are essentially 
confirmed by numerical simulations [7,8]. The finite coherence length 
$\lc$ -- that is: $x_i(n) \simeq \Lap{x}(n)$ for $i < \lc$, while 
for $i>\lc$ the $x_i$ are spatially irregular -- is due to numerical 
noise. Indeed $\lc$ decreases logarithmically with the noise level 
in the numerical simulations.

The open boundary condition can be seen as a defect in the chain. 
This scenario is essentially unchanged if one uses periodic boundary 
conditions and ``softer'' defects, such as, {\it e.g.}, the interchange of 
$\gamma_1$ and $\gamma_2$ in a finite fraction of sites [8].

The above argument neglects the fluctuations of the chaotic degree 
along the trajectory, since the Lyapunov exponent $\lambda_0$ gives 
only the typical chaotic degree. If one considers a temporal window
$[t-\Delta t/2,t+\Delta t/2]$ it is possible to repeat the arguments of 
Ref. [7] simply replacing $\lambda_0$ in equation (4) with 
the effective Lyapunov exponent [9] which measures the local 
exponential rate of growth 
for the tangent vector ${\bf z}$ around the time $t$:
$$ \chi_{\Delta t}(t) = {1\over \Delta t}
   \ln { |{\bf z}(t+\Delta t/2)|\over |{\bf z}(t-\Delta t/2)| }
  \eqno(5).$$
We expect that the scenario discussed in Ref. [7] may fail if the 
fluctuations of $\chi_{\Delta t}$ are strong. 
It is hence interesting to inquire what 
happens in the extreme cases. 

A rather natural way to address this problem is to study the 
system (1)-(2) with the single map 
   $$f(x) = x + c x^z \quad \hbox{\rm mod }1 \qquad (z\geq 1).
     \eqno(6)$$

For $z\ge 2$ the dynamical system $x(n+1)=f(x(n))$ shows 
a sporadic behaviour [10],{\it i.e.}
an initial disturbance grows in time as a stretched exponential with 
exponent less that one. The Lypunov exponent $\lambda_0$ hence vanishes.

In the case of $1\leq z<2$ the map (6) behaves as an ordinary chaotic 
system with positive Lyapunov exponent $\lambda_0$. Here we obtain 
for the system (1) the scenario discussed in Refs. [7,8] for 
the logistic map. We thus find 
that for symmetric couplings the chain does not synchronize, while in the 
asymmetric case a finite coherence length may appear depending on the value
of $\gamma_1-\gamma_2$ and $\lambda_0$. To be more specific we can 
distinguish two cases. In the first case $\rho_0>1$ and $|\rho(k)|<1$
-- {\it e.g.} when $z=1.5$, for which $\lambda_0=0.56$, $\gamma_1=0.7$ and 
$\gamma_2=0.01$ -- 
so that all the non-uniform eigenmodes are stable an the system 
synchronizes with a finite $\lc$ [7,8]. 
In the second case 
-- {\it e.g.} when $z=1.5$, $\gamma_1=0.7$ and $\gamma_2=0.1$ -- 
there exist some $k$ for which $|\rho(k)|>1$, and 
the system does not synchronize.

We now turn to the case of $z\ge 2$, when  the single map 
is sporadic. The results we shall report are obtained from numerical 
simulations with $\gamma_1=0.7$ and $\gamma_2$ which varies in the range 
$(0,\gamma_1]$. The typical sistem size considered is $N=200$, but 
we checked some results also for larger values of $N$. 

If the couplings are symmetric the systems is not spatially 
synchronized, $\lc\simeq 1$. Moreover, even if the Lyapunov exponent of 
the single map 
$\lambda_0$ is zero, the Lyapunov exponent $\lambda$ of the global system 
is always positive confirming the non relevance of the uniform mode for
the dynamics. 

As for the case of a chaotic map $f(x)$, when the couplings are 
not equal two qualitative different behaviours appear depending on the 
value of $\gamma_1-\gamma_2$. However, unlike the previous case, here the 
synchronization, when present, is complete, {\it i.e.} $\lc=N$. 
In Fig.1 we show $\tau$ -- the fraction of the time the system is 
completely synchronized -- and $\lambda$ -- the Lyapunov exponent -- 
as functions of $\gamma_2$. 

A first regime, with $\tau\simeq 1$, appears for $\gamma_1-\gamma_2$ 
large, which with our 
parameters means $\gamma_2\leq 0.2$. Here $\rho_0=1$ and $|\rho(k)|$ is 
well smaller than one for all $k$, {\it e.g.}, $\max_k |\rho(k)|=0.85$ for 
$\gamma_2=0.2$.
According to the above argument the uniform solution is locally 
stable. Indeed we find that the system synchronizes with $\lc=N$. 
A detailed analysis of the evolution reveals that the system 
spends most of the time completely synchronized. 
The ordered state is interrupted by intervals where the
coherence length becomes very small. This is a very 
different qualitative behaviour with respect to the case $z<2$ and the
case discussed in Refs. [7,8]. In fact in these latter cases the system 
is always partially synchronized, with $\lc$ undergoing small 
fluctuations about its time average.
All the simulations in this range of parameters 
gives a value of $\tau$ very close to one. For example, when
$\gamma_2 = 0.1$ we find, on a run of $5\times 10^6$ steps, 
$\tau=0.997$, while for $\gamma_2=0.2$, and on a run of 
$4.7\times 10^8$ steps, we find $\tau = 0.999$.
The Lyapunov exponent is always very small. It is difficult to decide if 
$\lambda$ is really going to zero or stays very small. This
difficulty is due to the fact that every time the system desynchronizes, 
the local Lyapunov exponent increases. As a consequence 
the numerical value of $\lambda$ does not saturate in time but has kicks
which increase its value, see Fig.2. 
We note that this incertitude on the value of $\lambda$ is directly
connected to that on $\tau$.

A different behaviour appears for $\gamma_1-\gamma_2$ small which, with our 
parameters, means $\gamma_2\geq 0.35$. In this regime $\rho_0=1$ but 
$|\rho(k)|$ is larger than one for some values of $k$, {\it e.g.}, 
when $\gamma_2=0.35$ we have $\max_k |\rho(k)|=1.04$. Thus the uniform 
state is locally unstable. Indeed we find that
the coherence length is almost always 
$\lc\simeq 1$ and the system is never completely synchronized -- 
at least on a run of $3\times 10^7$ steps. Moreover $\tau$ is zero and 
$\lambda$ is positive.
For example, $\lambda=0.74$ when $\gamma_2=0.35$.

These two behaviours appear to be related in a continuous way through the 
crossover region where $\max_k |\rho(k)|$ is very close to one, for example
$\max_k |\rho(k)|=0.92$ for $\gamma_2=0.3$. In this region 
$\lambda$ begins to be definitely non zero and
$\tau$ becomes smaller than one. For example, $\lambda\simeq 0.05$ and 
$\tau\simeq 0.9$ again for $\gamma_2=0.3$ on a run of 
$3\times 10^8$ time steps.

In order to stress the correlation between desynchronization and 
temporal chaoticity, we show in Fig.3 the effective Lyapunov 
exponent $\chi_{\Delta t}(t)$ and $\lc (t)$ as functions of $t$. 
It is evident that a large $\chi_{\Delta t}(t)$ corresponds to 
$\lc << N$ and {\it viceversa}. This happens both for small and 
large values of $\gamma_1-\gamma_2$.

We stress that since the map $f(x)$ is sporadic, and hence does not 
posses an asymptotic measure, the initial condition is relevant. 
In fact while the qualitative behaviour described above does non change,
at least for the values of $\gamma_2$ we used, 
the numerical value of $\lambda$ and $\tau$ my depend on the chosen 
initial condition.

This dependence is more dramatic in the crossover region, and makes
difficult to define clear boundaries. 

We conclude with some remarks. The behaviours we observed in the 
sporadic case are rather different from those one previously 
studied [7,8]: 

\noindent $a)$ one has a complete synchronization, {\it i.e.}
$\lc=N$;

\noindent $b)$ this complete synchronization does not change 
by the addition of a small noise, the effect of the noise 
is a small decreasing of the fraction, $\tau$, of the time 
in which $\lc=N$;

\noindent $c)$ there is a strong correlation between the spatial and 
temporal behaviour, in particular one observes that the 
effective Lyapunov exponent $\chi_{\Delta t}(t)$ is large in the 
intervals of time in which $\lc (t) << N$, while
$\chi_{\Delta t}(t) \simeq 0$ if $\lc (t)=N$.
 
\references
\numrefbk{[1]}{Kuramoto Y. 1984}%
              {Chemical Oscillations, Waves and Turbulence}%
              {(Springer-Verlag, New-York)}
\numrefbk{[2]}{Crutchfield J P and Kaneko K 1987}%
              {Directions in chaos}%
              {ed. Hao B L (Singapore: World Scientific) p 272}
\numrefbk{   }{Kaneko K 1990}%
              {Formation, Dynamics and Statistics of Patterns}%
              {ed.s Kawasaky K, Onnky A and Suzuki M 
               (Singapore: World Scientific)}
\numrefjl{[3]}{Gray C M, Konig P, Engel A K and Singer W 1989}%
              {Nature}{338}{334}
\numrefjl{   }{Eckhorn R {\it et al.} 1988}%
              {Biol. Cybern.}{60}{121}
\numrefjl{   }{Skarda C A and Freeman W J 1987}%
              {Behaviour. Brain Sci.}{10}{161}
\numrefjl{[4]}{Benzi R, Paladin G, Patarnello S, Santangelo P and 
               Vulpiani A 1986}%
               {J. Phys. A}{19}{3771}
\numrefjl{   }{Carnevale G F, McWilliams J C, Pomeau Y, Weiss J B and 
               Young W R 1991}%
               {Phys. Rev. Lett.}{66}{2735}
\numrefjl{[5]}{Leith C E 1971}%
              {J. Atmos. Sci.}{28}{145}              
\numrefjl{[6]}{Jensen M H 1989}%
              {\PRL}{62}{1361} 
\numrefjl{   }{Jensen M H 1989}%
              {Physica D}{32}{203} 
\numrefjl{[7]}{Aranson I, Golomb D and Sompolinsky H 1992}{\PRL}{68}%
              {3494} 
\numrefjl{[8]}{Biferale L, Crisanti A, Falcioni M and Vulpiani A 1993}%
              {J. Phys. A}{26}{L923}             
\numrefjl{[9]}{Paladin G and Vulpiani A 1987}%
              {Phys. Rep.}{156}{147}              
\numrefjl{[10]}{Gaspard P and Wang X-J 1988}%
              {Proc. Natl. Acad. Sci. USA}{85}{4591}

\figures
 
\figcaption{ Lyapunov exponent $\lambda$ (crosses) and probability of 
             complete synchronization $\tau$ (squares) {\it vs} $\gamma_2$.
             $N=200$ and $\gamma_1=0.7$. The integration steps
             are $3\cdot 10^8$ for $\gamma_2=0.3$, $4.7\cdot 10^8$ 
             for $\gamma_2=0.2$ and a few millions of 
             steps in the other cases.}           

\figcaption{Time evolution of the mean value defining the 
            Lyapunov exponent $\lambda$, for $\gamma_2=0.2$ 
            (lower curve) and $\gamma_2=0.3$.}

\figcaption{Effective Lyapunov exponent $\chi_{\Delta t}$
            and coherence length $\lc$ {\it vs} $t$. $N=200$, 
            $\gamma_1=0.7$, $\gamma_2=0.3$, $\Delta t=3\cdot 10^4$.} 

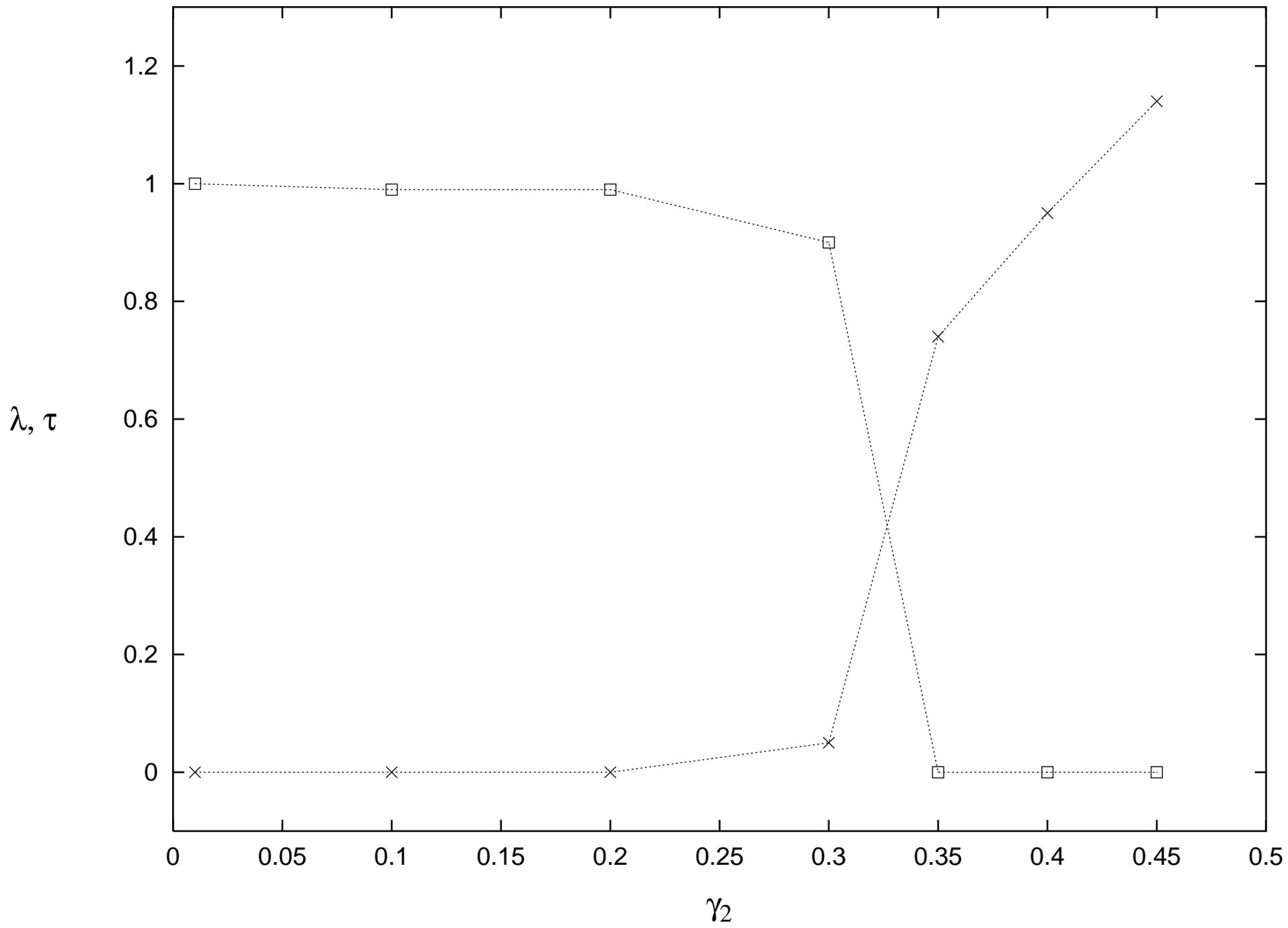
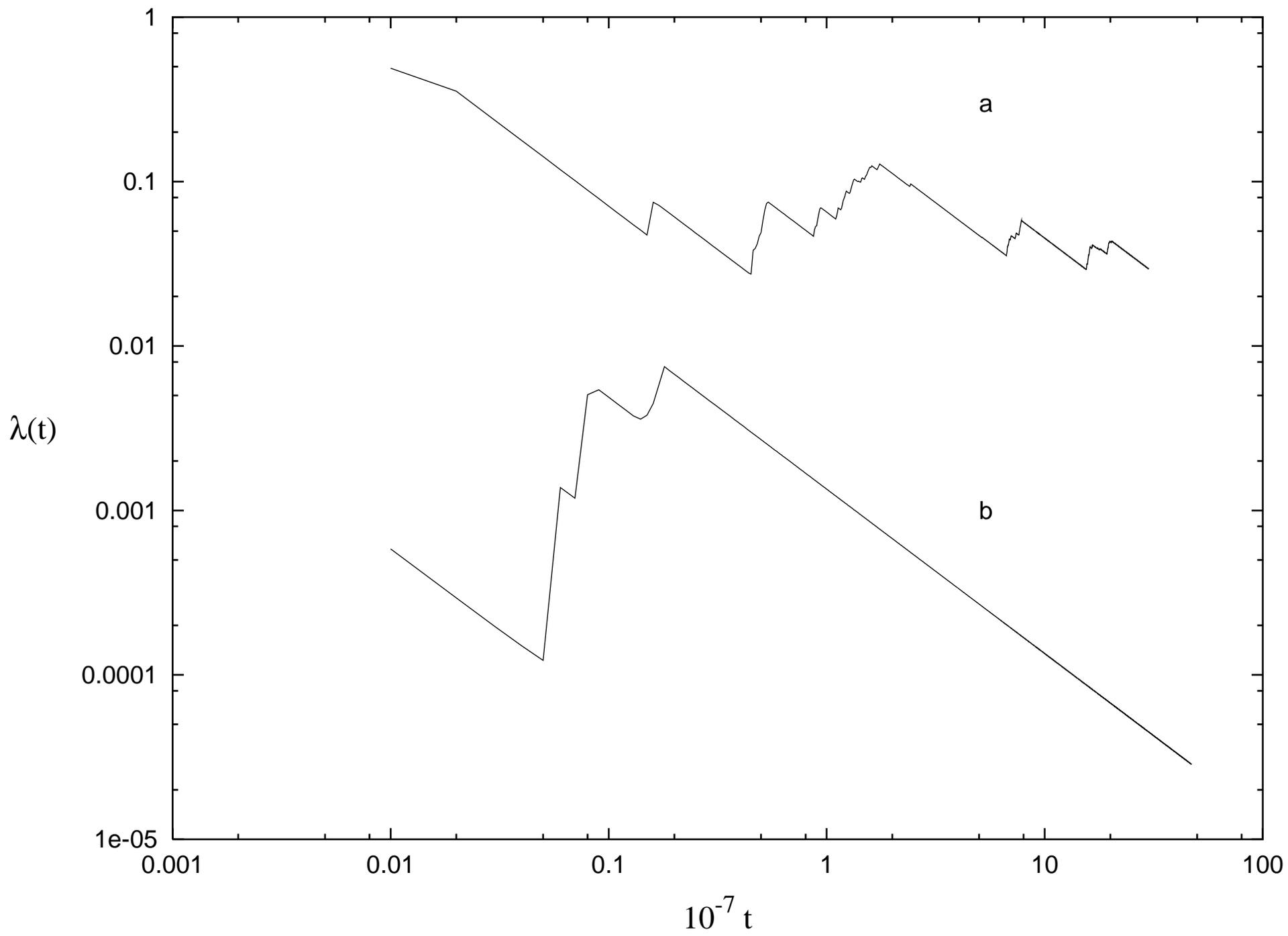
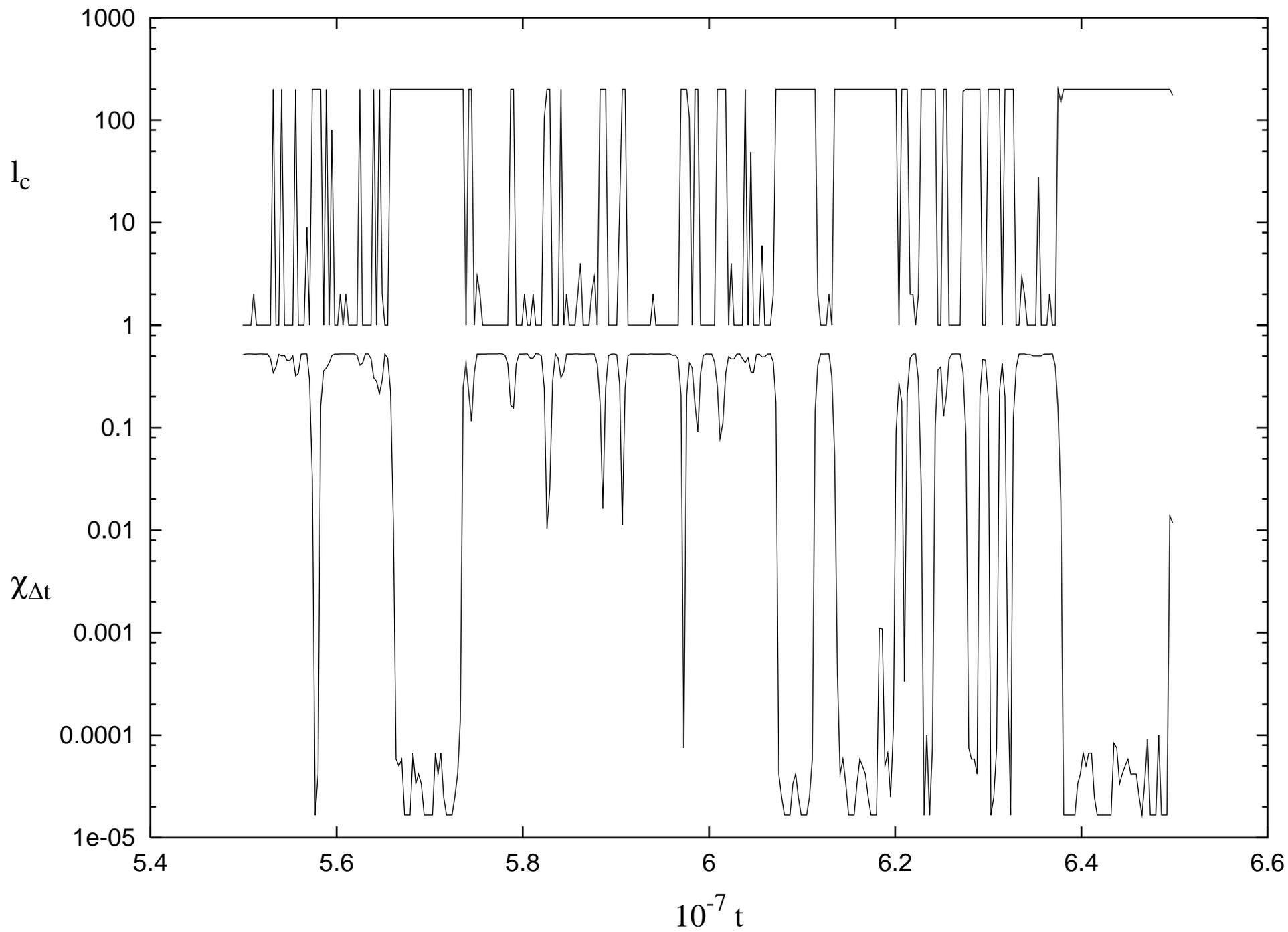
\bye